\documentclass[english,prl, twocolumn, footinbib, showpacs, showkeys]{revtex4}
\usepackage[T1]{fontenc}
\usepackage{color}
\usepackage{babel}
\usepackage{amsmath}
\usepackage{amssymb}
\usepackage[unicode=true,
 bookmarks=true,bookmarksnumbered=false,bookmarksopen=false,
 breaklinks=true,pdfborder={0 0 1},backref=false,colorlinks=true]
 {hyperref}
\hypersetup{
 pdfauthor={François Konschelle},
 citecolor=blue,linkcolor=magenta,urlcolor=blue}
\usepackage{breakurl}

\makeatletter
\@ifundefined{textcolor}{}
{%
 \definecolor{BLACK}{gray}{0}
 \definecolor{WHITE}{gray}{1}
 \definecolor{RED}{rgb}{1,0,0}
 \definecolor{GREEN}{rgb}{0,1,0}
 \definecolor{BLUE}{rgb}{0,0,1}
 \definecolor{CYAN}{cmyk}{1,0,0,0}
 \definecolor{MAGENTA}{cmyk}{0,1,0,0}
 \definecolor{YELLOW}{cmyk}{0,0,1,0}
}

\makeatother

\begin{document}

\title{Electromotive interference in a mechanically oscillating superconductor:
generalized Josephson relations and self-sustained oscillations of
a torsional SQUID}

\author{Fran\c{c}ois Konschelle}

\affiliation{Kavli Institute of Nanoscience - Delft University of Technology -
Lorentzweg 1 2628 CJ Delft - The Netherlands}

\email{francois_konschelle001@ehu.es}

\altaffiliation{Present address: Centro de F\'{i}sica de Materiales (CFM-MPC), Centro Mixto CSIC-UPV/EHU, Manuel de Lardizabal 4, E-20018 San Sebasti\'{a}n, Spain}

\author{Ya. M. Blanter}

\affiliation{Kavli Institute of Nanoscience - Delft University of Technology -
Lorentzweg 1 2628 CJ Delft - The Netherlands}

\date{\today}
\begin{abstract}
We consider the superconducting phase in a moving superconductor and
show that it depends on the displacement flux. Generalized constitutive
relations between the phase of a superconducting interference device
(SQUID) and the position of the oscillating loop are then established.
In particular, we show that the Josephson current and voltage depend
on both the SQUID position and velocity. The two proposed relativistic
corrections to the Josephson relations come from the macroscopic displacement
of a quantum condensate according to the (non-inertial) Galilean covariance
of the Schr\"{o}dinger equation, and the kinematic displacement of
the quasi-classical interfering path. In particular, we propose an
alternative demonstration for the London rotating superconductor effect
(also known as the London momentum) using the covariance properties
of the Schr\"{o}dinger equation. As an illustration, we show how these
electromotive effects can induce self-sustained oscillations of a
torsional SQUID, when the entire loop oscillates due to an applied
dc-current.
\end{abstract}



\maketitle
The interest in nano-mechanical systems dramatically increased recently.
For instance, the superposition of quantum states of a mechanical
resonator has been demonstrated \cite{O'Connell2010,Teufel2009},
realizing an important step towards the generation of mechanically
quantum dressed state at the mesoscopic level. These dressed states
open wide possibilities for using mechanical resonators for quantum-limited
detection and for quantum information. There are currently several
routes being explored towards these applications, based on the coupling
between mechanical resonators and either electrical, optical, or magnetic
systems. 

One of the route is to use a mechanical resonator embedded into a
superconducting quantum interference device (SQUID), see \emph{e.g.}
\cite{Zhou2006,Buks2006,Nation2008,Pugnetti2009} and references therein.
The first experiment demonstrated the possibility of the detection
of the resonance frequency and the quality factor of the resonator
by measurements of the voltage generated across the SQUID \cite{Etaki2008}.
Magnetic flux through the SQUID and the bias current served as two
control parameters. Later experiments \cite{Poot2010} investigated
back-action of the SQUID exerted on the mechanical resonator and found
qualitative agreement with the results of the theoretical modeling.
The experiments on the new generation suspended structure, a torsional
SQUID, demonstrated that the back-action may be so strong that the
SQUID enters the regime of self-sustained oscillations \cite{Etaki2013}.
It remained unclear, however, where such a large coupling between
the superconducting and mechanical degrees of freedom comes from.

So far, the description of the experimental setup ignored possible
electromotive effects of quantum nature. For instance, it is well
known that a cold quantum gas exhibits vortex states under rotation,
in an intuitive analogy with the Abrikosov lattice \cite{Bloch2007a}.
Nevertheless, the analogy is incomplete, because the Abrikosov lattice
is of electromagnetic nature, whereas the circulation vortices are
of mechanical origin. A situation when both these effects may compete
is precisely the situation when a quantum condensate made of charged
particles is mechanically displaced. Then, a superconducting condensate
may magnify electromotive effects when put under displacement. This
is illustrated by the striking Meissner or London momentum effects
\cite{b.london}. In short, the Meissner effect corresponds to the
generation of a displacement current which screens an applied magnetic
field, whereas the London momentum effect corresponds to the generation
of a macroscopic magnetic field which screens the displacement current
generated by the mechanical rotation of a superconductor. This surprising
effect can only exist when electromagnetism and mechanical displacement
compete together, and can be seen as the destruction of the mechanically
induced vortex lattice by the generation of an electromagnetically
induced lattice.

Another explanation of this effect lies in the well-known London theory
of the electrodynamics of superconductors, which corresponds to the
addition of an inertial term (proportional to the vector potential
$\mathbf{A}\propto\mathbf{j}$) into the otherwise viscous expressions
for circuit electromagnetism (\emph{i.e.} Ohm's theory). Because of
this inertial correction, the London theory is unable to take into
account a change of the inertial frame, in the sense that there is
no explicit need to specify in which inertial frame the superconductor
is supposed to be. Nevertheless, when a normal metal is attached to
a superconducting one, the normal electronic flow must be recovered
at the interface. Then, London proposed to correct his theory by imposing
a superconducting current to lag behind the lattice one, creating
a magnetic moment by virtue of the Amp\`{e}re law. To test the validity
of this retarded contribution, London designed the very simple experiment
of the rotating superconductor, which predicts the generation of a
macroscopic magnetic field induced by the rotation of a superconducting
sphere \cite{b.london}. The London's prediction was soon after verified
for both bulk \cite{Hildebrandt1964} and proximity effect \cite{Zimmerman1965}
systems, and latter for high-temperature superconductors \cite{Verheijen1990}
including heavy-fermion compounds \cite{Sanzari1996}.

More recently, the London expression for the inertial current was
considered in the framework of an effective and classical theory of
elastic superconductors, with prediction of a precise acoustic sensor
using Josephson systems, elasto-magnetic coupling between the motion
of the superconductor and the internal magnetic moment it produces,
in addition to some interesting effects in type-II superconductors,
see \emph{e.g.} \cite{Zhou1998} and references therein. It also continues
to attract some fundamental interests, being at the cornerstone between
mechanics and electromagnetism \cite{DeWitt1966,Liu1998,Anandan1999,Fischer2001,Tajmar2006}.

In this paper, we consider possible electromotive effects of quantum
mechanical origin in a moving superconductor. The results of this
study are twofold. In the first part, we will show that the London
momentum naturally appears in the context of Galilean covariance of
the Schr\"{o}dinger equation. Generalizing this demonstration to the
non-inertial case, we establish some generalized Josephson relations
in a moving superconducting circuit. In the second part, we will derive
constitutive relations linking the superconductor motion to the current
and voltage. Subsequently, we apply the arguments to the setup of
a suspended SQUID. As a simple illustration of the theory, we will
describe the regime when a static current can induce self-sustained
oscillations in a torsional SQUID, as shown in \cite{Etaki2013}.

We start with the correspondence of the London momentum and the Galilean
covariance of the Schr\"{o}dinger equation, leading to the generalization
of the Josephson relations when electromotive effects are taken into
account. We have in mind a superconducting circuit interrupted by
some Josephson junctions and when a region of the circuit is mechanically
oscillating. Then a part of the circuit is in the laboratory frame,
whereas the oscillating part is in a moving frame. In order to obtain
the electromotive contributions in a general situation, we constraint
the phase of the superconducting, macroscopic wave function to be
continuous all along the circuit. When the circuit forms a loop, and
when this loop is pierced by a magnetic flux $\Phi=\oint\mathbf{A}\cdot d\mathbf{l}$,
the phase continuity implies that the so-called gauge covariant phase
$\gamma=\varphi_{0}-2\pi\Phi/\Phi_{0}$ enters in the expressions,
where $\varphi_{0}$ is the initial condition phase \cite{b.tinkham}.

In any situation when part of the circuitry is moving, this gauge
covariant phase fails to describe the passage from the circuit at
rest to the moving part, possibly destroying the phase continuity
all along the circuit. From general properties of quantum mechanics,
the displacement of a particle generates specific phase factors. For
a massive particle, there are two possible phase factors which can
be added to the displaced wave-function. They correspond to the energy
correction $e^{\mathbf{i}Et/\hslash}\sim e^{\mathbf{i}mv^{2}t/2\hslash}$
due to the kinetic energy the particle acquires under displacement
and responsible for time dependent interference effects ; and the
displacement operator of the wave function $e^{\mathbf{i}px/\hslash}\sim e^{\mathbf{i}mvx/\hslash}$
responsible for position dependent interferences. 

Let us quantify this idea. The non-inertial transformation is defined
as the change from the unprimed laboratory-frame to the primed moving-frame
defined by $\mathbf{x}^{\prime}=\mathbf{x}-\mathbf{r}\left(t\right)$
and $t^{\prime}=t$. Then, the differential operators appearing in
the Schr\"{o}dinger equation transform according to $\partial/\partial t^{\prime}=\partial/\partial t+\mathbf{v}\left(t\right)\mathbf{\cdot\nabla}$
and\ $\mathbf{\nabla}^{\prime}=\mathbf{\nabla}$ where the velocity
$\mathbf{v}\left(t\right)=\mathbf{\dot{r}}$ is time dependent. Moreover,
one can show that the Schr\"{o}dinger equation for particle of mass
$m$ and charge $q$ in the laboratory frame, 
\begin{equation}
\mathbf{i}\hslash\dfrac{\partial\Psi}{\partial t}=\left[U+qV-\dfrac{\hslash^{2}}{2m}\left(\mathbf{\nabla-i}\dfrac{q}{\hslash}\mathbf{A}\right)^{2}\right]\Psi\left(\mathbf{x},t\right)\ ,
\end{equation}
is formally equivalent to the one with electromagnetic potentials
$V$ and $\mathbf{A}$, potential $U$ and wave function $\Psi$ expressed
in the displaced (primed) system with the correspondence laws (full
details of the calculations can be found in \emph{e.g.} \cite{Brown1999,Takagi1991},
see also note %
\footnote{Actually Brown and Holland discuss the \textit{inertial} Galilean
covariance of the Schr\"{o}dinger equation when an electromagnetic
field is taken into account \cite{Brown1999}, whereas Takagi discusses
the non-inertial covariance (\textit{i.e.} when the velocity is time-dependent
in the Galilean transformation $\mathbf{x}^{\prime}=\mathbf{x}-\mathbf{r}\left(t\right)$)
of the Schr\"{o}dinger equation \textit{without} external fields \cite{Takagi1991}.
To fuse these two results is straightforward, resulting in the covariance
rules \eqref{EQ_NG_covariance}-\eqref{EQ_NG_covariance_Psi} when
a non-inertial Galilean transformation is applied to the Schr\"{o}dinger
equation in the presence of electromagnetic fields.%
}) 
\begin{equation}
\left\{ \begin{array}{l}
U^{\prime}\left(\mathbf{x}^{\prime},t^{\prime}\right)=U\left(\mathbf{x},t\right)+m\mathbf{\dot{v}}\left(t\right)\\
V^{\prime}\left(\mathbf{x}^{\prime},t^{\prime}\right)=V\left(\mathbf{x},t\right)-\mathbf{v\cdot A}\left(\mathbf{x},t\right)\\
\mathbf{A}^{\prime}\left(\mathbf{x}^{\prime},t^{\prime}\right)=\mathbf{A}\left(\mathbf{x},t\right)
\end{array}\right.\label{EQ_NG_covariance}
\end{equation}
for displaced potentials and 
\begin{equation}
\Psi^{\prime}\left(\mathbf{x}^{\prime},t^{\prime}\right)=\exp\left[\dfrac{\mathbf{i}}{\hslash}\left(\int_{t_{0}}^{t}\dfrac{mv^{2}}{2}dt-m\mathbf{v\cdot x}\right)\right]\Psi\left(\mathbf{x},t\right)\label{EQ_NG_covariance_Psi}
\end{equation}
for the wave function ($t_{0}$ being the initial time when $\mathbf{v}\left(t_{0}\right)=\mathbf{0}$).
The simplest case of the Galilean transformation \cite{Brown1999}
with $\mathbf{\dot{v}}=\mathbf{0}$ is the obvious limit $\Psi^{\prime}=\exp\left[\mathbf{i}\left(Et-\mathbf{x\cdot p}\right)/\hslash\right]\Psi\sim\exp\left[\mathbf{i}\left(m/\hslash\right)\left(v^{2}t/2-\mathbf{x\cdot v}\right)\right]\Psi$
of the transformation \eqref{EQ_NG_covariance}-\eqref{EQ_NG_covariance_Psi}.
In addition the case of a rotating frame can be discussed with a time-dependent
velocity $\mathbf{v}\left(t\right)$ as well (more precisely a time-dependent
angular-momentum, see \cite{Takagi1991,Fischer2001} for more details),
and leads to similar covariance laws \eqref{EQ_NG_covariance}-\eqref{EQ_NG_covariance_Psi}.

From the transformation law \eqref{EQ_NG_covariance_Psi}, one can
show that the electromagnetic sources $\rho=q\left\vert \Psi\right\vert ^{2}$
and $\mathbf{j}=q\hslash$ Im$\left\{ \Psi^{\ast}\mathbf{\nabla}\Psi\right\} /m$
transform as (cf. \cite{Brown1999}) 
\begin{equation}
\left\{ \begin{array}{l}
\rho^{\prime}\left(\mathbf{x}^{\prime},t^{\prime}\right)=\rho\left(\mathbf{x},t\right)\\
\mathbf{j}^{\prime}\left(\mathbf{x}^{\prime},t^{\prime}\right)=\mathbf{j}\left(\mathbf{x},t\right)-\rho\left(\mathbf{x},t\right)\mathbf{v}\left(t\right)
\end{array}\right..\label{EQ_NG_covariance_sources}
\end{equation}
Here, the transformation law for the current is nothing else than
the London current substitution when a superconductor is displaced
\cite{b.london}. In addition to its justification using conservation
of current \cite{b.london}, thermodynamic properties \cite{Liu1998}
or equivalence principle \cite{DeWitt1966,Anandan1999}, we have thus
shown that the London substitution for the current can be substantiated
using the covariance of the Schr\"{o}dinger equation at the simple
Galilean relativity level. We even found that the London current is
still a correct expression for non-inertial displacements. We remark
that our argument is just the quantum version of the London's original
one, because phase continuity and current conservation are related
to each other.

Note that the full transformations \eqref{EQ_NG_covariance}, \eqref{EQ_NG_covariance_Psi}
and \eqref{EQ_NG_covariance_sources} are compatible with the usual
electromagnetic gauge transformations, and that the space-time transformation
laws for the electromagnetic potentials (and subsequently for the
electromagnetic fields) correspond to the \emph{magnetic-Galilean-limit}
when the electric displacement current does not exist, see \cite{LeBellac1973,Brown1999}
and note %
\footnote{In the so-called magnetic limit, the Galilean-covariant equations
of electromagnetic fields $\mathbf{E}$ and $\mathbf{B}$ are the
usual ones $\partial\mathbf{B}/\partial t+\mathbf{\nabla\times E}=\mathbf{0}$
and $\mathbf{\nabla\cdot B}=0$, whereas the $\mathbf{H}$ and $\mathbf{D}$
fields fail to verify the charge conservation: $\mathbf{\nabla\times H}=\mathbf{j}$
and $\mathbf{\nabla\cdot D}=\rho$. Nevertheless, one still verifies
$\mathbf{B}=\mathbf{\nabla\times A}$ and $\mathbf{E}=-\mathbf{\nabla}V-\partial\mathbf{A}/\partial t$,
which guaranty gauge invariance of the fields. Injecting $\mathbf{\tilde{A}=A}-\mathbf{\nabla}\xi$
and $\tilde{V}=V+\partial\xi/\partial t$ for any field $\xi\left(\mathbf{x},t\right)$
in $\left(\ref{EQ_NG_covariance}\right)$ will not modify the covariance
of the potentials transformation laws: then the covariance laws $\left(\ref{EQ_NG_covariance}\right)$
and the gauge invariance of the fields are compatible. Along \cite{Brown1999},
we would like to reinforce the surprising result that the covariance
laws of the Schr\"{o}dinger equations follow the \textit{magnetic}
limit \eqref{EQ_NG_covariance} of the so-called Galilean electromagnetism
for the potentials whereas the sources transform alongside the \textit{electric}
limit of Galilean electromagnetism, these two limits being incompatible
with each other in the Galilean relativity. So in short the London
momentum effect (given by the transformation \eqref{EQ_NG_covariance_sources})
follows from the covariance rules of quantum mechanics, and would
be difficult (if not impossible) to demonstrate using the Galilean
covariance of the classical dynamics. The price to pay for our demonstration
here is to stick with a pure time-dependent velocity $\mathbf{v}\left(t\right)$
(see after \eqref{EQ_quantum_electromotive} for the reason). The
reader is referred to the pedagogical articles \cite{LeBellac1973,Brown1999}
for more details about the Galilean electromagnetism and its covariant
transformations in the electric and magnetic limits.%
}.

In addition to the generation of the London momentum in \eqref{EQ_NG_covariance_sources},
the covariance laws \eqref{EQ_NG_covariance}-\eqref{EQ_NG_covariance_Psi}
induce interference effects in coherent circuits. For instance, in
order for the phase to be continuous all along a displaced superconducting
path, one must include the kinematic terms \eqref{EQ_NG_covariance_Psi}
into the gauge-covariant phase definition. The generalized gauge-covariant
phase is thus 
\begin{equation}
\varphi=\varphi_{0}-\dfrac{2\pi}{\Phi_{0}}\oint\mathbf{A\cdot}d\mathbf{l}+\dfrac{2\pi}{\Phi_{0}}\dfrac{m_{e}}{e}\oint\mathbf{v\cdot}d\mathbf{l}\label{EQ_EmE_SC}
\end{equation}
where $2m_{e}$ is the Cooper pair mass, and the phase (difference)
is defined as $\varphi=\int\mathbf{\nabla}\varphi\cdot d\mathbf{l}$.
Obviously, the velocity integral in Eq.\eqref{EQ_EmE_SC} disappears
when the superconducting condensate remains at rest along the path,
and thus it connects the displacement velocity to the phase difference
in the Josephson system and eventually to the Josephson current. This
extra contribution would also disappear if there were not two distinct
frames moving relative to each other with the velocity $\mathbf{v}$.
For instance, the two frames of reference can be a superconductor
(as the moving frame) and a measuring apparatus (say a magnetometer
playing the role of the laboratory frame), like for the rotating superconducting
effect \cite{b.london}. 

The inclusion of the relativistic correction in \eqref{EQ_EmE_SC}
suggests that some subsequent contributions will be associated to
the so-called second Josephson relation. Indeed, the second Josephson
relation connects the time derivative of the superconducting phase
(difference) with the energy difference across the weak-links, \emph{i.e.}
to the electromagnetic work the superconducting charges undergo when
traveling across the junction. The total electromagnetic work is found
according to the complete Lorentz force. Using the generic relation
\cite{Flanders2010} 
\begin{equation}
\dfrac{d}{dt}\iint\mathbf{B\cdot}d\mathbf{S}=\iint\left[\dfrac{\partial\mathbf{B}}{\partial t}-\mathbf{\nabla\times v\times B}+\mathbf{v}\left(\mathbf{\nabla\cdot B}\right)\right]\mathbf{\cdot}d\mathbf{S}\label{EQ_flux_Flanders}
\end{equation}
for any $\mathbf{B}$ and $\mathbf{S}$ fields, when the infinitesimal
surface element $d\mathbf{S}$ is time dependent, with $\mathbf{v}\left(t\right)$
the velocity field of the contour, one obtains 
\begin{equation}
\dfrac{d\varphi}{dt}=\dfrac{2\pi}{\Phi_{0}}\dfrac{m_{e}}{e}\dfrac{d}{dt}\oint\mathbf{v\cdot}d\mathbf{l}+\dfrac{2\pi}{\Phi_{0}}\oint\left(\mathbf{E}+\mathbf{v\times B}\right)\mathbf{\cdot}d\mathbf{l}\label{EQ_quantum_electromotive}
\end{equation}
for the phase-voltage relation, using Faraday's law. For $\mathbf{v}=\mathbf{0}$,
we recover the (second) Josephson relation $\dot{\varphi}=2eV/\hslash$
\cite{b.tinkham}. In addition to the first term, which generates
electromagnetic fields due to the motion of accelerating charges,
the last term in Eq.\eqref{EQ_quantum_electromotive} is another electromotive
one, caused by the moving surface the magnetic flux threads. Thus,
even a static magnetic field can generate such a retardation effect,
due to the kinematic displacement of the interfering paths in the
presence of an electromagnetic field. 

The two expressions \eqref{EQ_EmE_SC} and \eqref{EQ_quantum_electromotive}
and their explicit derivations are our first results. We believe they
are rather general, because the notation $\varphi=\int\mathbf{\nabla}\varphi\cdot d\mathbf{l}$
above was a generic phase difference along a superconducting path,
eventually a closed one. Because their derivation used the covariance
properties of the Schr\"{o}dinger equation under a non-inertial transformation,
Eqs.\eqref{EQ_EmE_SC} and \eqref{EQ_quantum_electromotive} describe
any situation when phase coherence is preserved at large distance
between two sub-systems moving relatively to each other with velocity
$\mathbf{v}\left(t\right)$ (after proper identification of the elementary
mass and charge at works in some specific examples, which may not
necessarily be the Cooper pair ones). Obviously, superconductivity
exhibits such a possibility, and we will focus on superconducting
systems in the following. 

In order to apply relations \eqref{EQ_EmE_SC} and \eqref{EQ_quantum_electromotive}
to explicit circuits, we remark that Eqs.\eqref{EQ_NG_covariance}
and \eqref{EQ_NG_covariance_Psi} are presumably invalid for a true
velocity field $\mathbf{v}\left(\mathbf{x},t\right)$: the covariance
of the Schr\"{o}dinger equation would be destroyed if $\mathbf{r}$
was a function of both time and position, because of the appearance
of mixing derivatives \cite{Takagi1991}. In addition, the fact that
the velocity explicitly depends in time only was crucial to obtain
the London expression \eqref{EQ_NG_covariance_sources} for the current
(recall that $\mathbf{j}$ is defined as a gradient interference).
Nevertheless, we note that only some space integrals of the velocity
appear in \eqref{EQ_EmE_SC} and \eqref{EQ_quantum_electromotive},
which supposedly signifies that a weak spatial dependency of $\mathbf{v}\left(\mathbf{x},t\right)$
(say with some characteristic magnitude in space smaller than the
size of the contour of the integral) would not alter the generalized
Josephson relations found here (see also \cite{Anandan1999,Fischer2001}
when alternative demonstrations of \eqref{EQ_EmE_SC} are given for
a generic $\mathbf{v}\left(\mathbf{x},t\right)$). Due to their integral
representations, we also remark that only \textit{global} properties
of the velocity field appear in the expressions for the phase difference
\eqref{EQ_EmE_SC} and its time derivative \eqref{EQ_quantum_electromotive}. 

We now apply the generalized Josephson relations \eqref{EQ_EmE_SC}
and \eqref{EQ_quantum_electromotive} to the SQUID geometry, see \textit{e.g.}
\cite{Etaki2008,Poot2010,Etaki2013}. More precisely, we discuss the
problem of an \textit{entirely suspended} SQUID, the so-called torsional
SQUID, when the entire loop is oscillating and pierced by a magnetic
flux, as realized recently \cite{Etaki2013}.

For simplicity, we suppose the entire loop to behave as a simple harmonic
oscillator of mass $m$, length $\ell$, quality factor $Q$, and
resonance frequency $\omega_{0}$, according to
\begin{equation}
\ddot{x}+\dfrac{\omega_{0}}{Q}\dot{x}+\omega_{0}^{2}x=g\left(i+j\right)\label{EQ_x}
\end{equation}
where $g\approx\ell BI_{0}/m$ is a geometric acceleration, such that
$g\left(i+j\right)$ represents the Lorentz force acting on the oscillator
when $B$ is the external magnetic field, $\left(i+j\right)$ being
the total current through the loop, see \eqref{EQ_RCSJ_EM}. The overdot
refers to total time derivative. 

Let us now make plausible the appearance of a non-trivial electromotive
effect for torsional SQUID. If only a part of the loop was oscillating,
the contribution $\oint\mathbf{v\cdot}d\mathbf{l}\approx0$ in \eqref{EQ_EmE_SC}
would vanish, since the velocity of the oscillator would mainly be
orthogonal to the loop in that case. Nevertheless, in a torsional
SQUID, this relation becomes non-local since the entire loop is oscillating,
and the usual relation $\oint\mathbf{v\cdot}d\mathbf{l}=\iint\left(\nabla\times\mathbf{v}\right)\cdot d\mathbf{S}$
applies, with $\mathbf{S}$ the surface-vector of the entire loop.
Due to the non-connectedness of the ring geometry, such a term does
\textit{not} vanish in elastic media \cite{Muskhelishvili2010,Hackl1988}.
This involved theorem of differential geometry can be picturesquely
described in the following way: the torsional SQUID exhibits torsional
elastic modes, which in turn create some global vorticity $\nabla\times\mathbf{v}$
around the loop, and thus generate a non-vanishing electromotive contribution
in \eqref{EQ_EmE_SC}. More precisely, this global vorticity generates
a pseudo-angular-momentum to the elastic loop. Then one can use the
correspondence between a pure time-dependent and non-inertial Galilean
transformation (as used to obtain \eqref{EQ_NG_covariance_sources},
\eqref{EQ_EmE_SC}, and \eqref{EQ_quantum_electromotive}) and the
non-inertial transformation towards a rotating frame (see \textit{e.g.}
\cite{Takagi1991}) to ensure that the contribution $\oint\mathbf{v\cdot}d\mathbf{l}\neq0$
is non-trivial for a torsional SQUID. In the following, we suppose
$\oint\mathbf{v}\cdot d\mathbf{l}\approx\ell\dot{x}$ for notational
simplicity.

The non-mechanical SQUID is characterized by two degrees of freedom,
which we take to be the sum and the difference of the phases of the
junctions, respectively $\varphi_{\pm}$ \cite{b.tinkham}. When the
SQUID is mechanically oscillating, these two phases are modified due
to the electromotive effects in \eqref{EQ_EmE_SC} and \eqref{EQ_quantum_electromotive}.
First, the phase difference reads then
\begin{equation}
\varphi_{-}=-\varphi_{e}-\kappa_{-}x+\dfrac{k}{\omega_{0}}\dot{x}-\beta j\ ,\label{EQ_superconducting_electromotive_effect}
\end{equation}
where $\beta=4\pi LI_{0}/\Phi_{0}$ is the self-inductance of the
loop, $\varphi_{e}=\pi\Phi/\Phi_{0}$, $\Phi$ the external flux across
the loop and $\Phi_{0}=\pi\hbar/e$ the superconducting flux quantum,
$\kappa_{-}\approx\pi B\ell/\Phi_{0}$ is a flux-to-phase ratio, and
$k\approx\left(m_{e}\omega_{0}/\hslash\right)\ell$ within our approximation.
Note that the extra contribution is just the usual phase factor $\exp\left[\mathbf{ip\cdot x}\right]$
corresponding to the extra path length due to the displacement of
the mechanical oscillator, when the impulsion $\mathbf{p}\approx m\mathbf{v}/\hslash$
is related to the velocity of the mechanical loop, the length of which
is $\ell$. Also, $k$ is the inverse of an effective Compton wavelength
\cite{Zimmerman1965}, related to the inertia of the electrons because
of their lag behind the mechanical oscillations of the lattice. 

Comparing the two mechanical contributions in \eqref{EQ_superconducting_electromotive_effect}
leads to the ratio $B/\omega_{0}\approx m_{e}/2e\approx3\text{ng/C}$
which seems extremely small. Nevertheless, being a velocity term,
the Compton contribution has to be compared with the $Q$-factor of
the oscillator, which may eventually be large enough to make interesting
relativistic effects observable at the nanoscale. Also, it is interesting
to mention that the $k$ term in \eqref{EQ_superconducting_electromotive_effect}
survives in the absence of magnetic field. Thus, the observation of
this electromotive effect in the absence of external magnetic field
should be a clear demonstration that a non-trivial displacement of
charges generates a complete electromagnetic field at the quantum
level.

The phase difference is not affected by the contribution \eqref{EQ_quantum_electromotive},
and the $\mathbf{v\times B}$ term affects only the phase sum $\varphi_{+}$.
Eq.\eqref{EQ_quantum_electromotive} implies that the voltage in a
torsional SQUID is defined as 
\begin{equation}
V=\dfrac{\Phi_{0}}{2\pi}\dfrac{d}{dt}\left[\varphi_{+}+\kappa_{+}x-\dfrac{k}{\omega_{0}}\dot{x}\right]\label{EQ_quantum_electromotive_effect}
\end{equation}
with $\kappa_{+}\approx\kappa_{-}\approx\pi B\ell/\Phi_{0}$ in our
approximation. Expression \eqref{EQ_quantum_electromotive_effect}
appears natural: because the position of the oscillating loop is equivalent
to a flux for the SQUID, its velocity $\dot{x}$ generates a voltage.
The Compton $k$-term plays the role of a usual Bremsstrahlung in
\eqref{EQ_quantum_electromotive}. 

With the help of \eqref{EQ_superconducting_electromotive_effect}
and \eqref{EQ_quantum_electromotive_effect}, the equations of motion
for the SQUID can be easily written in the quasi-classical approximation
(also called RCSJ-model \cite{b.tinkham}),
\begin{equation}
\left\{ \begin{array}{l}
i=\sin\varphi_{+}\cos\varphi_{-}+\dfrac{\dot{\varphi}_{+}+\kappa_{+}\dot{x}-k\ddot{x}/\omega_{0}}{\omega_{c}}\\
\;\;\;\;\;\;\;\;\;\;\;\;\;\;+\dfrac{RC}{\omega_{c}}\left(\ddot{\varphi}_{+}+\kappa_{+}\ddot{x}-\dfrac{k}{\omega_{0}}\dddot{x}\right)\\
j=\sin\varphi_{-}\cos\varphi_{+}+\dfrac{\dot{\varphi}_{-}}{\omega_{c}}+\dfrac{RC}{\omega_{c}}\ddot{\varphi}_{-}
\end{array}\right.\label{EQ_RCSJ_EM}
\end{equation}
with $\omega_{c}=2\pi RI_{0}/\Phi_{0}$ the characteristic frequency
of the SQUID, $I_{0}$ the critical current of each Josephson junction,
$R$ and $C$ the resistance and capacitance of the shunted Josephson
junctions, $i=I/2I_{0}$ and $j=J/2I_{0}$, where $I$ and $J$ are
the bias and self-circulating current, respectively.

Eqs.\eqref{EQ_x}-\eqref{EQ_RCSJ_EM} represent the set of non-linear
coupled differential equations of a torsional SQUID. The limit $\kappa_{\pm}\rightarrow0$
and $k\rightarrow0$ in \eqref{EQ_RCSJ_EM} corresponds to the usual
SQUID description without any electromotive effect \cite{b.tinkham}.
The equations \eqref{EQ_x}-\eqref{EQ_RCSJ_EM} are extremely hard
to solve. Nevertheless, one can easily show that they lead to self-sustained
oscillations of the loop induced by a static current.

Indeed, in the over-damped limit and without self inductance (\emph{i.e.
}$RC/\omega_{c}\rightarrow0$ and $\beta\rightarrow0$), the macroscopic
time averaged voltage can be approximated as \cite{b.tinkham} 
\begin{equation}
\dfrac{\left\langle V\right\rangle }{RI_{0}}=\sqrt{\left(\dfrac{I}{2I_{0}}\right)^{2}-\cos^{2}\left(\varphi_{e}+\kappa_{-}x-\dfrac{k}{\omega_{0}}\dot{x}\right)}\label{EQ_voltage}
\end{equation}
in the limit $\omega_{0}/\omega_{c}\ll1$. This limit is realized
in actual experiments \cite{Etaki2008,Poot2010,Etaki2013}, when $\omega_{0}/\omega_{c}\lesssim\left(10^{-3}-10^{-4}\right)$.
Then, in the harmonic equation of motion \eqref{EQ_x}, one has to
realize that the Lorentz force is present only in the resistive branch
of the SQUID response, and Eq.\eqref{EQ_x} can be approximated as
\begin{equation}
\ddot{x}+\dfrac{\omega_{0}}{Q}\dot{x}+\omega_{0}^{2}x\approx g\dfrac{\left\langle V\right\rangle }{RI_{0}}\label{EQ_self-sustained}
\end{equation}
which exhibits self-sustained oscillations when $gk/\omega_{0}^{2}\gtrsim Q^{-1}$
or in our approximation $\ell^{2}BI_{0}Q/m\omega_{0}\gtrsim\hslash/m_{e}$,
the quantum of circulation. Thus, to apply a static current to a torsional
SQUID can generate self-sustained mechanical oscillations, as already
demonstrated in \cite{Etaki2013}. Here, we present an alternative
explanation based on electromotive interferences in the mechanical
loop, instead of the capacitive coupling already discussed in \cite{Etaki2013}.
Note that the equations of motion in both cases are similar in the
limit $k/\omega_{0}\ll1$ in \eqref{EQ_voltage}, which make the origin
of the self-sustained oscillations demonstrated in \cite{Etaki2013}
a bit unclear. Nevertheless, for the experiment \cite{Etaki2013},
the capacitive coupling seems unable to generate the self-sustained
oscillation by a few orders of magnitude, whereas the condition $gk/\omega_{0}^{2}\gtrsim Q^{-1}$
is verified.

In conclusion, we established and discussed electromotive phenomena
in the Josephson physics, based on the gauge-covariance of the Schr\"{o}dinger
equation under a non-inertial Galilean transformation. It leads to
both an alternative demonstration of the London momentum \eqref{EQ_NG_covariance_sources},
and to generalized Josephson relations taking into account the displacement
of the superconductor, see \eqref{EQ_EmE_SC} and \eqref{EQ_quantum_electromotive}.
We then apply these generic expressions to the case of a torsional
SQUID. When only a part of the SQUID is oscillating, there is in general
no electromotive effect associated with the displacement. In contrary,
when the entire loop is oscillating as in the torsional SQUID, we
established some generalized differential equations describing the
coupled evolution of the mechanical and charge degrees of freedom,
see \eqref{EQ_quantum_electromotive}-\eqref{EQ_RCSJ_EM}. In addition,
the electromotive effects can lead to self-sustained oscillations
of the torsional SQUID. The electromotive effects we discussed may
be relevant for the description of the measurement of the motion of
a quasi-classic elastic loop, and/or for the description of the back-actions
a SQUID exerts onto itself when the entire loop is elastic. More precisely,
the Compton-like $k$-term seems to be important for back-action effect,
because it generates magnetic flux ; whereas the kinetic interference
$\kappa_{+}$-term transforms existing magnetic field into voltage,
and might be important for detection purpose when the voltage is monitored. 
\begin{acknowledgments}
We thank I. \textsc{Bouchoule}, D. \textsc{Dulin, }S. \textsc{Etaki},
G. \textsc{Rousseaux, }R.P. \textsc{Tiwari }and\textsc{ }H.S.J. \textsc{van
der Zant} for stimulating discussions, and B. \textsc{Bergeret} for
bibliographic facilities. We acknowledge the financial support of
the Future and Emerging Technologies program of the European Commission,
under the FET-Open project QNEMS (233992) and of the Netherlands Foundation
for Fundamental Research on Matter (FOM).
\end{acknowledgments}

\bibliographystyle{apsrev4-1}
\bibliography{electromotive}

\end{document}